\documentclass[usletter]{article}

\usepackage{INTERSPEECH2016}

\usepackage{graphicx}
\usepackage{amssymb,amsmath,bm}
\usepackage{textcomp}
\usepackage{balance}
\usepackage{color}

\ninept

\title{Speaker Sincerity Detection based on Covariance Feature Vectors and Ensemble Methods}

\makeatletter
\def\name#1{\gdef\@name{#1\\}}
\makeatother \name{{\em Mohammed Senoussaoui$^1$, Patrick Cardinal$^1$, Najim Dehak$^2$, Alessandro L. Koerich$^1$ }}

\address{$^1$\'{E}cole de Technologie Sup\'{e}rieure, Montr\'{e}al, Canada \\
  $^2$Johns Hopkins University, Baltimore, USA \\
  {\small \tt mohammed.senoussaoui.1@ens.etsmtl.ca, patrick.cardinal@etsmtl.ca} \\
  {\small \tt ndehak3@jhu.edu, alessandro.koerich@etsmtl.ca}
}

\begin{document}
\maketitle
\begin{abstract}
Automatic measuring of speaker sincerity degree is a novel research problem in computational paralinguistics. This paper proposes covariance-based feature vectors to model speech and ensembles of support vector regressors to estimate the degree of sincerity of a speaker. The elements of each covariance vector are pairwise statistics between the short-term feature components. These features are used alone as well as in combination with the ComParE acoustic feature set. The experimental results on the development set of the Sincerity Speech Corpus using a cross-validation procedure have shown an 8.1\% relative improvement in the Spearman's correlation coefficient over the baseline system.

\end{abstract}


\section{Introduction}
\label{sec:intro}
Human speech carries emotional information that can be transmitted through a change in vocal parameters \cite{Scherer2003}. In particular, the task of measuring the degree of sincerity of a speaker is very challenging and complex even if the literal meaning of his elocution is perfectly understood. Humans usually use distinct vocal cues to express their sincerity, which often makes the perception of sincerity from the speaking style easier than from the literal contents \cite{Eriksson2010, BJDP:BJDP2062, Rigoulot201448}.

Computation paralinguistics is a relatively recent research field and its main purpose is to change the classical interaction way between human beings and machines by embedding into it some of the humane natural capacities \cite{schuller2013computational}. Since 2009 a periodic challenge on Computational Paralinguistics (ComParE) have been organized and the degree of perceived sincerity of speakers is one of the three challenges proposed for the 2016 ComParE Challenge \cite{Schuller20111062, schullerInterSpeech2016}. The participants are challenged to estimate the degree of perceived sincerity of 32 speakers from a total of 911 instances. Each instance was rated in terms of perceived sincerity using a rating scale ranging from 0 (not sincere at all) to 4 (extremely sincere) by at least 13 annotators.     

From the purely theoretical standpoint, adopting powerful features able to model most faithfully the signal is helpful for the subsequent process of estimating the degree of perceived sincerity. Besides the baseline feature set resulting from the computation of various functionals over low-level descriptor contours \cite{schullerInterSpeech2016}, in this paper we investigate the suitability of two different levels of features for estimating the degree of perceived sincerity: low-level features given by the short-term features; high-level features given by fixed length vectors based on pairwise statistics calculated from the low-level features. As short-term features, we adopted the well-known Mel Frequency Cepstral Coefficients (MFCC) augmented by their Shifted Delta Cepstral (SDC) \cite{sdc2002} in order to model the temporal aspect of the speech signal. For the sake of fixed length feature vectors that are more suitable for most of the classification and regression models, usually, some individual statistics (e.g. mean, variance, min, max, etc) are extracted from the short-term variable length features. Instead of using only these simple statistics, another option is to employ pairwise statistics (e.g. covariance, correlation, etc) that model the relationship between the individual features since such statistics are able to describe better the useful information embedded in the short-term features. In this paper, the covariance based features \cite{Tuzel2006} are used as high-level features to train regression models. Furthermore, we also evaluate the combination of regression models trained on different feature sets to explore any eventual complementary between high-level and low-level features.          

This paper is organized as follows. Section \ref{sec:Featuredescription} reviews the different signal representations adopted in this work. Section \ref{sec:LearningAlgorithms} describes the regression models. The experimental results on the Sincerity Speech Corpus dataset are presented and discussed in Section\ref{sec:Experimentations}. Finally, conclusions as well as some future directions are presented in the last two sections section.

\begin{figure*}[!ht]
\includegraphics[scale=0.8]{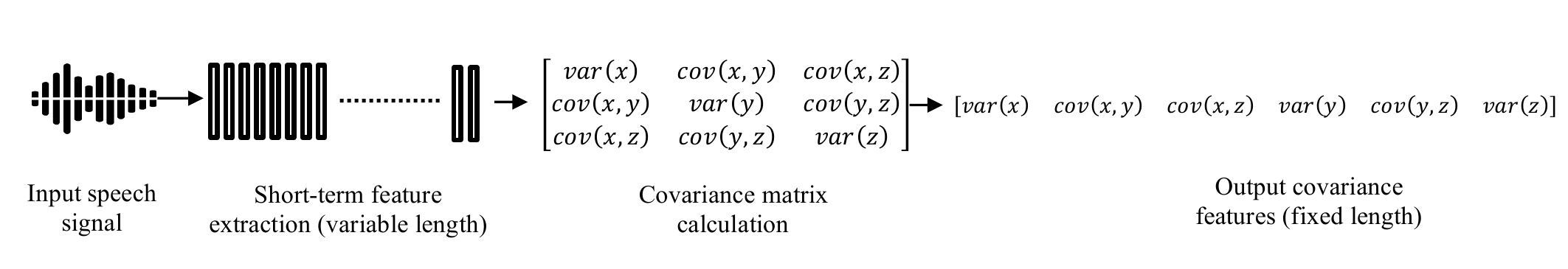}
\caption{{\it An overview of the covariance-based feature extraction from a given speech signal.}}
\label{fig:CovFea}
\end{figure*}

\section{Feature description}
\label{sec:Featuredescription}
The official ComParE baseline feature vectors contain 6,373 components representing various functionals calculated over low-level descriptor (LLD) contours which cover descriptors used in  speech processing, music information retrieval, and general sound analysis \cite{Scherer2003}. A detailed description of these features can be found in \cite{Scherer2003, Weninger13}.

Even if the baseline feature vectors cover a broad set of descriptors, some descriptors that have been recently proposed, such as i-vectors, covariance features and augmented covariance features, have not been exploited to this particular task of estimating the degree of perceived sincerity of speakers. Therefore, we investigate if such descriptors may provide diverse and/or complementary information relative to the baseline features that allows to improve the performance measures.

\subsection{I-Vectors}
The speech representation using the i-vectors became the state-of-the-art in the text-independent speaker recognition field since 2009 \cite{Dehak10frontend}. An i-vector is a utterance-based vector with two main advantages over conventional MFCCs: ability to keep a wide range of useful information using a moderate dimensionality (in the range of few hundreds). One limitation of the i-vector representation is the need of a relatively large amount of training data as well as a slightly complicated training procedure.

The basic idea behind the i-vector representation can be summarized into two main steps: i) based on the maximum {\it a posteriori} (MAP) adaptation we model the short-term sequence of features by adapting the mean vectors of a well trained Gaussian Mixture Model (GMM) base; ii) reduce the dimensionality of the concatenated mean vectors (named supervectors) of the resulting utterance-dependent GMM using a factor analysis model.

As we aforementioned, the training of an i-vector extractor requires a relative large dataset, which is not the case of the development dataset provided for the sincerity prediction task of the ComParE 2016 challenge. On the other hand, we have already achieved a good performance in the AVEC2014 Challenge for the task of depression level prediction using the TIMIT dataset \cite{senoussaoui2014model}. Such a dataset is comprised of 6 300 recordings of 630 different speakers, each recording is around 3 seconds duration, recorded using 16 bits precision and a sampling rate of 16 kHz. Therefore, we have trained our i-vector extractor using the TIMIT database.    

\subsection{Covariance Features (CF)}
The extraction of short-term features from the raw speech signal is a standard in most of speech processing based systems. However, the variable length of the short-term sequence representation usually requires a mapping into a fixed-length representation space in order to augment the flexibility of choosing a powerful classification or regression model. A current practice is to extract simple individual statistics (such as mean, variance, range, min, max, etc) to model the short-term feature distribution. Another way of extracting powerful fixed-length features is by considering the pairwise relationships between the individual features measured, for example, by the covariance, the correlation, etc. The covariance-based features were proposed for the first time for the task of object detection \cite{Tuzel2006} and they were further used in several works that deal with object detection and tracking \cite{Porikli2006, Tuzel2007}. Lately, covariance-based features have been successfully employed for emotion classification from speech \cite{Ye2008}. 

Figure \ref{fig:CovFea} illustrates the process of extracting covariance-based features from a speech signal. For a given input speech signal we first extract a sequence of short-term features. The sample covariance matrix is then computed for such a sequence of features and the covariance feature vector is obtained by vectorizing the upper (or lower) triangular part of the covariance matrix. Note that the dimension $D$ of the resulting CF vector is given as follow: 
\[ D = \frac{d\times(d+1)}{2} \]
where $d$ is the dimension of the short-term features. 

In this work, as we aforementioned, we have used as short-term features, the MFCCs augmented by their shifted delta cepstral (SDC) \cite{sdc2002}. The SDC features have been extracted using the Kaldi Toolkit \cite{Povey_ASRU2011}. The number of cepstral coefficient was 20, the distance between blocks was set to three, the size of delta advance and delay was one and the number of blocks in advance of each frame to be concatenated was set to seven. This configuration produces feature vectors of dimension 160. Another configuration including the pitch adds two more coefficients resulting in feature vectors of 176 dimensions.   

The extraction of these features is extremely easy and it does not need any extra data\footnote{This is not true in the case we need to reduce the dimensionality of the covariance features.} to train the extractor as is the case of i-vectors. These advantages make the covariance-based features suitable for many applications dealing with limited data or limited computational resources. 

\subsection{Augmented Covariance Features (ACF)}
Besides the variances, the covariance features adopt only the covariances as pairwise statistics However, as we have aforementioned, features based on individual statistics are also widely used to represent the speech signal. Therefore, we propose to augment the covariance features using some individual statistics in order to take advantage of both of features. In this work we have  augmented the covariance features using the mean, the skewness as well as the kurtosis individual statistics. Thus, the dimension $D^\prime$ of the resulting ACF vector is given as follows:  
\[ D^\prime = D+d \times 3 \]
where 3 denotes the number of different individual statistic added (i.e. mean, skewness and kurtosis).

\section{Learning Algorithms}
\label{sec:LearningAlgorithms}
Based on our previous work on emotion detection \cite{Cardinal2015} and native language detection \cite{Senoussaoui2016}, the machine learning paradigm chosen is based on kernel machines. This section briefly describes the learning algorithms used in this work. For all of them, the dlib library \cite{dlib:09} has been used.  

\begin{table*} [!ht]
\vspace{2mm}
\centerline{
\begin{tabular}{|l|c|c|c|c|c|c|}
\cline{2-7}
\multicolumn{1}{c|}{}  & \multicolumn{6}{|c|}{Spearman's correlation coefficient}\\ 
\cline{2-7}
\multicolumn{1}{c|}{}  & \multicolumn{2}{|c|}{SVR} & \multicolumn{2}{|c|}{RVM} & \multicolumn{2}{|c|}{KRR}\\ 
\hline
Features      & LOSO CV & Test Set & LOSO CV & Test Set & LOSO CV & Test Set\\
\hline
\hline
BF         		& 0.475 & -- & 0.458 & -- & 0.428 & --\\
i-vector                & 0.426 & -- & -- & -- & -- & --\\
CF            & \bf{0.514} & 0.451 & 0.458 & -- & 0.470 & --\\
CF with pitch & 0.505 & -- & 0.500 & -- & 0.458 & --\\
ACF             & 0.510 & -- & 0.486 & -- & 0.463 & --\\
ACF with pitch  & 0.512 & -- & \bf{0.503} & -- & \bf{0.471} & --\\
\hline
\end{tabular}}
\caption{\label{table:Base_results} {\it Spearman's correlation coefficient obtained with SVR, RVM and KRR algorithms on different feature vectors (BF: baseline features; CF: covariance features; ACF augmented covariance features).}}
\end{table*}

\subsection{Support Vector Regression}
Support vector regressors (SVR) with linear kernels have been shown to give competitive accuracy for problems where the number of features is very large as well as in problems where the number of features are the much larger than the number of instances. In such cases, a nonlinear mapping does not improve the performance. Linear SVRs are good enough and they enjoy much faster training/testing. Given a set of training vectors $\textbf{x}$ and the respective target values $\textbf{y}$, where $\textbf{x}\in\Re^n$ and $\textbf{y}\in\Re$, linear SVR finds a model $\textbf{w}$ such that $\textbf{w}^T\textbf{x}$ is close to the target value $\textbf{y}$. It solves the following regularized optimization problem.

\[ \min_{\textbf{w}} f(\textbf{w})\equiv\frac{1}{2}\textbf{w}^T\textbf{w}+C\sum\xi_\epsilon (\textbf{w};\textbf{x},\textbf{y}) \]
where $C>0$ is the regularization parameter and

\[ \xi_\epsilon (\textbf{w};\textbf{x},\textbf{y}) = \max(|\textbf{w}^T\textbf{x}-\textbf{y}|-\epsilon,0) \]
\noindent is the $\epsilon$-insensitive loss function associated with $(\textbf{x},\textbf{y})$.

\subsection{Relevance Vector Machine}
Relevance vector machines (RVM) has been proposed by Tipping to circumvent some inconveniences of conventional support vector machines, such as the requirement to estimate trade-off parameters and the need to use a "Mercer" kernel \cite{Tipping:1999}. As described in \cite{Huang:2015}, the goal of the RVM training is to learn the regression function:

\[ y(\textbf{x};\textbf{w}) = \textbf{w}^T\Phi(\textbf{x}) + \pmb{\epsilon} \]
where $\textbf{x}$ is the input feature vector, $\textbf{w}$ is an estimated set of sparse regression parameters, $\psi(\textbf{x}) = [\Phi_1(\textbf{x}),\Phi_2(\textbf{x}),...,\Phi_p(\textbf{x})]$ is the set of transforms performed on $\textbf{x}$ and $\pmb{\epsilon} = [\epsilon_1,\epsilon_2,...,\epsilon_N]^T$ is the set of training noise terms that are assumed to be Gaussian~$N(0,\sigma^2)$

RVM presents the learned regression model as the most relevant set of extracted feature dimensions, meaning that the technique performs both dimensionality reduction and feature selection. Considering the size of the baseline feature vector and the small amount of available data, this is an interesting characteristic \cite{Huang:2015}.

\subsection{Kernelized Ridge Regression}
The Kernelized ridge regression (KRR), described in \cite{Murphy:2012}, combines the ridge regression and the kernel trick. The ridge regression is a linear least square regression with a l2-norm regularization that encourages small $w$:
\[ \sum_{i=1}^N( (f(x_i) - y_i)^2 ) + \lambda \cdot dot(\textbf{w},\textbf{w}) \]
where $dot(x,y)$ is the dot product and $f(x) = dot(\textbf{x},\textbf{w}) -b$. The Kernel trick is applied by replacing the dot product by a kernel function:

\[ \sum_{i=1}^N( (f(x_i) - y_i)^2 ) + \lambda*K(\textbf{w},\textbf{w}) \]
where $K(\textbf{x},\textbf{y})$ is the kernel and $f(x) = K(\textbf{x},\textbf{w}) -b$. The fitting the KRR can be done in closed-form and is usually faster SVR. Another advantage is that there is no model hyper-parameters such as with the SVR and RVM algorithms.

\begin{figure*}[!htpb]
\centering
\includegraphics[width=17.5cm]{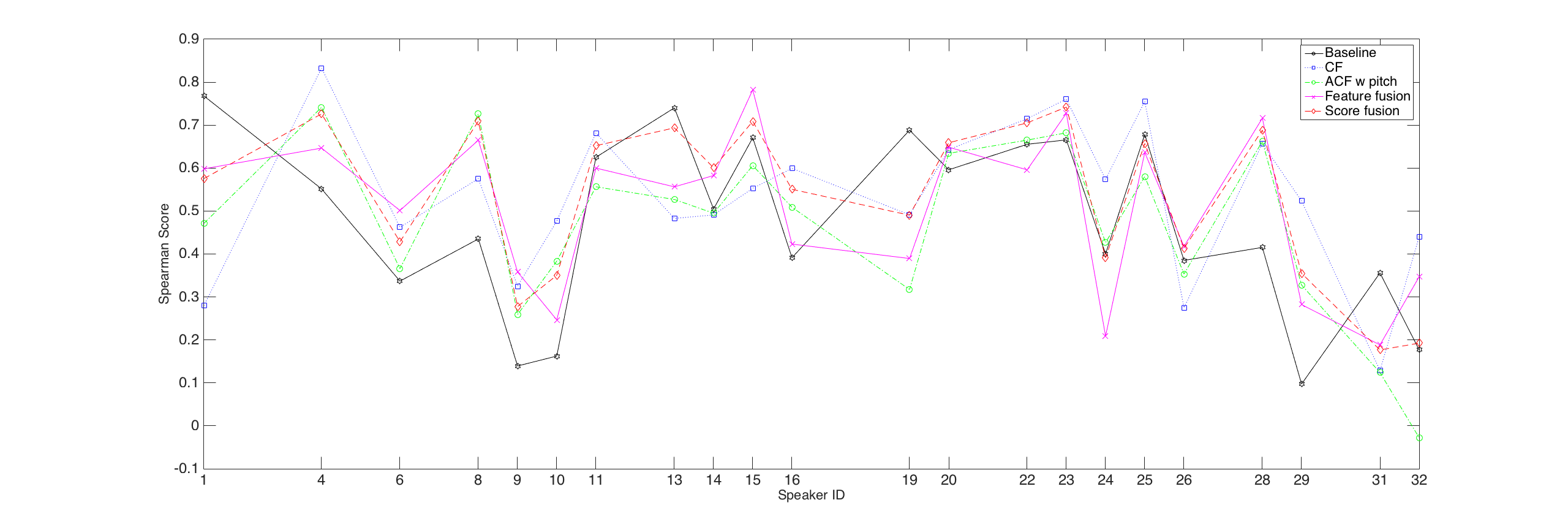}
\caption{{\it Spearman's correlation coefficient for each speaker considering different feature sets and fusion schemes.}}
\label{fig:curves}
\end{figure*}

\section{Experimental Results}
\label{sec:Experimentations}
The performance of the proposed approach was evaluated on the 2016 ComParE development and test datasets using a leave-one-speaker-out cross-validation (LOSO CV) procedure and a hold-out procedure respectively. This dataset is based on the Sincerity Speech Corpus provided by the Columbia University. A number of speakers were asked to read six different sentences, with each sentence read in four different prosodic styles \cite{schullerInterSpeech2016}. 

The full dataset contains 72 min of speech by 32 speakers and a total of 911 instances. Each instance is rated in terms of perceived sincerity using an ordinal rating scale from zero ( not sincere at all) to four (extremely sincere). The golden standard consists of the standardized ratings across at least 13 annotators. The development dataset is made up of 655 speech samples from 22 speakers while The test dataset is made up of 256 speech samples from the remaining speakers.  

Table \ref{table:Base_results} shows the results achieved using a support vector regressor (SVR) with linear kernel, a relevance vector machine (RVM) and a kernelized ridge regression (KRR) respectively. We can see that all the proposed feature vectors surpass the performance of the baseline features for all regression algorithms considering the development dataset (LOSO CV). The best result is achieved using the Covariance Features (CF) and it has a Spearman's correlation coefficient of 0.514. This configuration has been submitted to be evaluated on the test set and result was quite low with a correlation coefficient of 0.451 compared to the 0.602 achieved by the baseline. One hypothesis for this deceiving result on the test dataset was the very high dimension of the CF vectors (12,880). Therefore, we have evaluated the reduction of the dimensionality of the CF space as an attempt to remedy such a problem.

Given the extraction procedure of the CF features, the reduction of the dimensionality can be carried out in two different levels. First, we can reduce directly the dimensionality of the output space (i.e. the 12880-dimensional space of CF features), or instead, we can reduce the dimensionality of the input space (i.e. the D-dimensional space of MFCC features). Given the high dimensionality of the CF vectors and the lack of training data we have preferred to perform the dimensionality reduction in the input space using a Principal Component Analysis (PCA). This PCA was trained on a subset of about 5 hours taken from the test set of the ComParE 2016 Native Language dataset \cite{schullerInterSpeech2016}. The results on such reduced space of representation are shown in Table \ref{table:SVR_results}, where three different dimensions for the input feature space were evaluated. In this case, only the SVR was used because it was the approach that provided the best results in our previous experiments.  

\begin{table} [!ht]
\vspace{2mm}
\centerline{
\begin{tabular}{|l|c|c|c|}
\hline
Input space    & CF vector       & LOSO CV & Test set\\
dimension   &  dimension      & & \\
\hline
\hline
60     & 1 830    & 0.478 & -- \\
80     & 3 240    & 0.493 & 0.475 \\
100    & 5 050    & 0.481 & -- \\
160*   & 12 880   & 0.514 & 0.451 \\
\hline
\multicolumn{4}{l}{\scriptsize *Original input space dimension.}
\end{tabular}}
\caption{\label{table:SVR_results} {\it Spearman's correlation coefficient obtained with SVR on reduced feature vectors.}}
\end{table}

Besides evaluating the proposed feature sets individually, we have also observed a certain complementarity between them and the baseline features, as shown in Figure \ref{fig:curves}. Therefore, in an attempt to further improve the performance we have employed two strategies to combine these feature sets: a fusion at feature level which concatenates feature sets; fusion at score level that combines the output scores of the regression models using an average rule. Table \ref{table:FeatureFusion_results} shows that the fusion at both levels is able to further improve the performance over the proposed and the baseline feature sets.  

\begin{table} [!ht]
\vspace{2mm}
\centerline{
\begin{tabular}{|l|c|c|}
\cline{2-3}
\multicolumn{1}{c|}{}  & \multicolumn{2}{|c|}{Correlation}\\ 
\hline
Feature level      & LOSO CV & Test Set\\
\hline
\hline
BF + CF                 & 0.545 & -- \\
BF + CF + pitch         & 0.552 & 0.503 \\
BF + ACF                 & 0.541 & -- \\
BF + ACF + pitch         & 0.550 & -- \\
BF + CF + i-vector    & 0.545 & -- \\
BF + CF + pitch + i-vector  & \bf{0.553} & -- \\
BF + ACF + i-vector          & 0.541 & -- \\
BF + ACF + pitch + i-vector  & 0.550 & -- \\
BF + CF + ACF + pitch  & 0.537 & -- \\
\hline
Score level      & LOSO CV & Test Set\\
\hline
\hline
BF + CF                                  & 0.538 &  -- \\
BF + CF + ACF\_pitch                   & 0.561 &  -- \\
BF + CF + ACF\_pitch + Feature fusion  & \bf{0.562} & \bf{0.547} \\
\hline
\end{tabular}}
\caption{\label{table:FeatureFusion_results} {\it Spearman's correlation coefficient obtained with the fusion of feature sets at both feature and score level.}}
\end{table}


 
\section{Conclusions}
\label{sec:Conclusions}
This paper shows that feature sets that have been proposed for text-independent speaker recognition and for object recognition and later for emotion classification from speech, can be successfully used to detect speaker sincerity from speech.

These features are used alone as well as in combination with the ComParE acoustic feature set. The experimental results on the development set of the Sincerity Speech Corpus using a cross-validation procedure have shown 8.1\% to 18.1\% relative improvement in the Spearman's correlation coefficient over the baseline system considering the best individual feature set and the best fusion of feature sets respectively. Unfortunately we have not achieved the same improvement on the test dataset. Our best result was 0.547 against 0.602 reported in \cite{schullerInterSpeech2016}.

Nevertheless, these preliminary results open up several perspectives for our future work. Based on the differences in the performances among the several feature sets we have observed that we could achieve a Spearman's correlation coefficient as high as 0.740 using an oracle to select the best feature set for each speaker in the development dataset. Furthermore, we have also evaluated the similarity between the feature vectors presented in the test dataset and the speakers in the development dataset. As a result of such an analysis we have found out that, for example, 15\% of the test instances are similar to the instances of the speaker 19, as well as 16.4\% of the test instances are similar to the instances of the speakers 4 and 28. This suggests that a dynamic selection mechanism may help to further improve the performance.

 
\section{Acknowledgments}
The authors acknowledge funding from FRQNT and the CSAIL Lab for giving access to their cluster.


\clearpage
\balance
\eightpt

\end{document}